\documentclass{appolb}

\newcommand\del{\delta}
\newcommand\bea{\begin{eqnarray}}
\newcommand\eea{\end{eqnarray}}
\newcommand\Sf{{\rm S}_{1,2}}
\newcommand\SSS{{\rm S}}
\newcommand\Li{{\rm Li}}
\newcommand\EV{\, \mbox{\boldmath $E$}}
\newcommand\Ev{\, \mbox{\boldmath $E$}}
\newcommand\PV{\, \mbox{\boldmath $P$}}
\newcommand\Pv{\, \mbox{\boldmath $P$}}
\newcommand\Uv{\, \mbox{\boldmath $U$}}
\usepackage{epsfig}
\begin{document}
\begin{flushleft} DESY 02-099 \hfill {\tt hep-ph/0207259}\\
July 2002 \end{flushleft}

\vspace{4cm}
\begin{center}
{\bf \LARGE
Universal QED Corrections to Polarized\\

\vspace*{1mm}
Electron Scattering in Higher Orders}

\vspace{2.5cm}
{\large Johannes Bl\"umlein and Hiroyuki Kawamura}

\vspace{1.0cm}
{\large Deutsches Elektronen--Synchrotron, DESY\\ Platanenallee 6
D-15738 Zeuthen, Germany}

\vspace{\fill}
{\bf Abstract}
\end{center} \noindent
We derive QED radiators for the universal corrections to polarized
electron scattering. To 5th order in the coupling constant the
flavor non-singlet and singlet contributions are calculated. We derive
the non-singlet and singlet exponentiation of the leading terms
$\propto (\alpha \ln^2(x))^k$ to all orders.

\vspace{1cm} \begin{center}
{\sf Contribution to DIS 2002, April 2002, Cracow.} \end{center}

\title{Universal QED Corrections to Polarized\\
Electron Scattering in Higher Orders
\thanks {Talk presented by H. Kawamura.}%
}
\author{\it Johannes Bl\"umlein
and
Hiroyuki Kawamura
\address{\it Deutsches Elektronen--Synchrotron, DESY,\\
Platanenallee 6, D--15738 Zeuthen, Germany}
}
\maketitle

\begin{abstract}
\noindent
We derive QED radiators for the universal corrections to polarized
electron scattering. To 5th order in the coupling constant the
flavor non-singlet and singlet contributions are calculated. We derive
the non-singlet and singlet exponentiation of the leading terms
$\propto (\alpha \ln^2(x))^k$ to all orders.
\end{abstract}
\section{Introduction}
\noindent
QED corrections to differential cross sections can be quite large in
some kinematic regions both in deep inelastic scattering~\cite{RCDIS}
and $e^+e^-$ scattering. Besides of the process-dependent radiative
QED corrections in all the scattering processes classes of
{\sf universal contributions} emerge, which, when having been known
once at high precision have not to be calculated again, but can just
be used in novel applications. These contributions may be attributed to
radiation from outer legs (light fermions) and correspond to those
due to {\it i)} leading order mass factorization and  {\it ii)} the
resummation of leading order universal contributions $\propto (\alpha
\ln^2(x))^k$ to all orders. The latter terms form universal pieces
in the higher order anomalous dimension matrices~\cite{SX1}--\cite{SX4}.
In Ref.~\cite{BKA} we derived the corrections and give a brief summary
here.
\section{Complete Leading Order Solutions to $O(\alpha^5)$}
\noindent
Unlike the case for parton distributions in QCD the evolution equations
cannot be solved simply numerically since the sources are $\propto
\delta(1-x)$, the Mellin transforms of which show no damping behaviour
as {\sf Re}($N) \rightarrow - \infty$. Therefore the corresponding Mellin
convolutions have to be carried out analytically. All convolution
integrals to be used are given in \cite{BKA}. We
define non-singlet and singlet distribution functions for the electron,
$D_{\rm NS}(a,x) = D^{e^-}(a,x) - D^{e^+}(a,x),~~D_{\Sigma}^e(a,x) =
D^{e^-}(a,x) + D^{e^+}(a,x)$, with $a = \alpha/(4\pi)$. The singlet
distribution mixes with the photon distribution $D^\gamma(a,x)$ under
evolution.
\subsection{The Non-Singlet Case}
\noindent
We calculate the Mellin convolution in $x$ space term by term using the
Mellin transformation of  Nielsen and related
functions \cite{BKU,BKA}.
Using different techniques these contributions were also calculated in
\cite{PRZ,ARB}. A very
compact result can be obtained using both soft--photon exponentiation 
and the asymptotic solution at small 
$x$~\cite{JEZ} for all orders~:
\begin{eqnarray}
D_{\rm NS}(x,\beta) &=& \Biggl[\frac{\exp[\beta/2(3/4-\gamma_E)]}
{\Gamma(1+\beta/2)} \frac{\beta}{2} (1-x)^{\beta/2-1}
2\frac{{\rm I}_1\left((-\beta\ln(x))^{1/2}\right)}{[-\beta \ln(x)
]^{1/2}} \nonumber\\ & &~~~~~~~~~~~~
~~~~~~~~~~~~~~~~~~~~~~~~~~~~~~~\times
\sum_{n=0}^{\infty}    \left(\frac{\beta}{2}\right)^n \Psi_n(x)\Biggr]_+.
\end{eqnarray}
The functions $\Psi_k(x)$ are given by
\begin{eqnarray}
\Psi_0(x) &=& 1+x^2\\
\Psi_1(x) &=& -\frac{1}{2} \left[(1-x)^2+x^2\ln(x)\right]\\
\Psi_2(x) &=& \frac{1}{4} (1-x)\left[1-x-x\ln(x)+(1+x)\Li_2(1-x)\right]\\
\Psi_3(x) &=&  -\frac{1}{48} \Biggl\{6(1-x^2)[2 \Li_3(1-x)+ \Li_2(1-x)]
                          +5(x-1)^2 \nonumber\\ & &
                          + \frac{1}{12} x^2 \ln^3(x)
          + (1+7 x^2)[ \ln(x)\Li_2(1-x) + 2\Sf(1-x)]
          \nonumber\\ & &
          -\left(\frac{1}{2}+6 x - \frac{13}{2} x^2\right) \ln(x)
          \Biggr\} \\
\Psi_4(x) &=&  \frac{1}{96} \Biggl\{(1-x^2)
\Biggl[24 \Li_4(1-x)+ 12\Li_3(1-x)
               - \frac{5}{2} \SSS_{1,3}(1-x) 
               \nonumber\\ & &
               - 12 \SSS_{2,2}(1-x)
               - \frac{3}{2} \ln(x)\Sf(1-x)
              -\frac{1}{4} \ln^2(x) \Li_2(1-x)
                \Biggr] \nonumber  \\ & &
              + 4(1+x^2) \Li_2^2(1-x) + 7 \Li_2(1-x)
               \nonumber\\ & &
              + 2 (1+7x^2)  \ln(x) \Li_3(1-x)
              -\left(\frac{3}{4} + 5x -\frac{23}{4} x^2\right) 
              \ln(x) \nonumber\\ & &
              -\frac{1}{12}x(1-x) \ln^3(x)
              -\frac{1}{48}x^2 \ln^4(x)
              + (1-x)^2 \left[\frac{7}{2} 
              +\frac{1}{8}\ln^2(x) \right]  \nonumber \\
              & &
               +(1-8x+7x^2)\Biggl[\ln(x) \Li_2(1-x) +2\Sf(1-x)\Biggr]
              \Biggr\}~.
\end{eqnarray}
Here $\beta=(2/\pi)\int^s_{m_e^2} (ds'/s')\alpha(s^\prime)$.
These results agree with Ref.~\cite{PRZ}.
\subsection{The Singlet Case}
\noindent
Since there is no exponentiation formula in the singlet case, the
evolution equations cannot be reduced to a simple form as in the 
non-singlet case and term by term convolution forms the final result.

The singlet distribution for electrons is given in matrix form as,
\begin{equation}
{\bf D}_\Sigma(a,x)={\EV}_s(a,x) \otimes \left(
\begin{array}{c}
\del(1-x)\\
0
\end{array}
\right).
\end{equation}
Here $\otimes$ indicates both matrix multiplication and Mellin
convolution. ${\EV}_s(a,x)$ is the evolution operator given by 
\begin{eqnarray} 
\EV_s(x,\beta) &=& {\bf 1}~\delta(1-x)
+ \sum_{k=1}^\infty \frac{1}{k!}\PV_0^{(k)}(x) \left(-\frac{1}{\beta_0}
\ln\left(\frac{a}{a_0}\right)\right)^k
\\
\PV_0^{(k)}(x)&=&\otimes_{l=1}^k \PV_0~.
\end{eqnarray}
The Mellin convolutions can be separated into the non--singlet part
and a pure singlet part for the fermions. The expressions are rather
lengthly and are given in Ref.~\cite{BKA} in explicit form.
\section{Resummation of small $x$ logarithms}
\noindent
The resummation of the terms $\propto \alpha^n\log^{2n}(z)$ is carried 
out using infrared
evolution equations~\cite{IRE}. Applications to QCD
evolution were considered in \cite{SX1,SX2} and for unpolarized QED
in \cite{SX4}. The contributions under consideration form the most
singular parts of the anomalous dimensions in the respective order as
$x \rightarrow 0$ and have to be treated as such in the evolution 
equations.
\subsection{Non-Singlet Case}
\noindent
The leading double-log terms in non-singlet evolution kernel are given 
in Mellin space by~\footnote{Note that {\bf none} of these representations
exhibits poles in $N$ but only branch cuts.}
\begin{equation} 
{\cal M}[P_{\rm NS}, x  \rightarrow 0](N,a)
=\frac{N}{2}\left\{1-\sqrt{1+\frac{8a}{N^2}
\left[1-2\sqrt{1-\frac{8a}{N^2}}\right]}\right\},
\end{equation}
which are converted into $x$ space using serial representations. 
The solution is given by
\begin{equation}
D_{NS,x\rightarrow 0}(z)=\sum_{k=0}^\infty c_k \int^s_{m_e^2}\frac{dq^2}{q^2} 
a^{k+1}(q^2)\log^{2k}(z)~.
\end{equation}  
The coefficients $c_k$ agree with those of Ref.~\cite{SX4}.
\subsection{The Singlet Case}
\noindent
In the singlet case the small $x$ leading double log kernel in given by
\begin{eqnarray}
{\bf P}(x,a)_{x \rightarrow 0} = \sum_{l=0}^{\infty}
{\bf P}^{(l)}_{x \rightarrow 0} a^{l+1} \ln^{2l}(x) = \frac{1}{8\pi^2}
{\cal M}^{-1}\left[{\bf F}_0(N,a)\right](x)~,
\end{eqnarray}
where the matrix ${\bf F}_0(N,a)$ is determined by
\begin{eqnarray}
{\bf F}_0(N,a) &=& 16 \pi^2 \frac{a}{N} {\bf M}_0 - \frac{8a}{N^2}
{\bf F}_8(N,a){\bf G}_0 + \frac{1}{8\pi^2} \frac{1}{N} 
{\bf F}_0^2(N,a)~,\\
{\bf F}_8(N,a) &=& 4\pi^2\left(1-\sqrt{1-\frac{8a}{N^2}}\right)
{\bf M}_8~,
\end{eqnarray}
with the  matrices
\begin{eqnarray}
{\bf M}_0 = \left(\begin{array}{rr} 1 &  -2  \\ 2 & 0 \end{array}
\right),~~~~~~{\bf M}_8 = \left( \begin{array}{rr} 1 & - 1 \\
0 & 0 \end{array} \right),~~~~~~{\bf G}_0 = 
\left( \begin{array}{rr} 1 & 0 \\
0 & 0 \end{array} \right)~.
\end{eqnarray}
Solving these equations we obtain\footnote{Similar to a result
found for QCD in \cite{SX2}, the off diagonal elements of these matrices
are equal up to their sign.}
\begin{eqnarray}
{\bf P}_{z\rightarrow 0}^{(0)}= 
\left(\begin{array}{rr} 2 &  -4  \\ 4 & 0 \end{array} \right),~~~
{\bf P}_{z\rightarrow 0}^{(1)}= 
\left(\begin{array}{rr} 2 &  -4  \\ 4 & 0 \end{array} \right),~~~
\cdots
\end{eqnarray}
In the singlet case, the resummed evolution equation cannot easily be
solved analytically since the kernels in different orders do not commute
in general and one has to take into account a larger number of
terms.  However, unlike the case in the fixed--order iteration, the
Mellin transform of $\ln^{2k}(x)$ is suitably damped as {\sf Re}$(N)
\rightarrow -\infty$. The singlet solution is obtained in terms
of the \Uv--matrix formalism, see e.g. [4b]. The
\Uv-matrix is obtained order by order in Mellin space as,
\bea
[{\rm \Uv}_k(N),{\rm \Pv}_0(N)/\beta_0]={\rm \Pv}_k/\beta_0
+\sum_{i=1}^{k-1}{\rm \Pv}_i(N){\rm \Uv}_i(N)+k{\rm \Uv}_k(N)~,
\eea
where $\Pv_k$ are the most singular anomalous dimension matrices
$\propto \ln^{2k}(x)$. After having solved these equations the result
is used to perturb around $\Ev_s$

\bea
{\bf D}_s(a,N)=\left({\rm 1}+\sum^\infty_{k=0}a^k{\rm \Uv}_k(N)\right)
{\rm \Ev}_s(a,a_0,N)
\left({\rm 1}+\sum^\infty_{k=0}a^k_0{\rm \Uv}_k(N)\right)^{-1}~.
\eea 
\section{Summary}
\noindent
We calculated the universal QED corrections to hard scattering processes
due to light fermions for the non--singlet and singlet channels. The
leading order solution was evaluated up to  $O(\alpha^5)$ in analytic 
form. The universal contributions $\propto (\alpha \ln^{2}(x))^k$
are found in terms of analytic series expansions to all orders and
are combined with the foregoing corrections numerically.
More details of the calculation and results are given in Ref.~\cite{BKA}.

\end{document}